 \documentclass[final,3p,times]{elsarticle}
 \usepackage[colorlinks,linkcolor=blue,citecolor=green]{hyperref}

\usepackage{amssymb}
\usepackage{graphicx}
\usepackage{amsfonts}
\usepackage{subfigure}
\usepackage{amsthm}

\usepackage[T1]{fontenc} 
\usepackage{graphicx}
\usepackage{dcolumn}
\usepackage{amsfonts}

\usepackage{amssymb}
\usepackage{amsmath}
\usepackage{graphicx}
\usepackage{subfigure}
\usepackage{makeidx}
\usepackage{capt-of}
\usepackage{float}
\usepackage{placeins}
\usepackage{perpage}
\usepackage{mwe}
\usepackage{kantlipsum}
\graphicspath{{images/}}

\usepackage{graphicx,epsfig}
\usepackage{amsmath}
\usepackage{subfigure}
\usepackage{dcolumn}
\usepackage{amsmath}
\usepackage{amsfonts}
\usepackage{amssymb}
\usepackage{braket}

\usepackage{amssymb}
\usepackage{amsmath}
\usepackage{graphicx}
\usepackage{subfigure}
\usepackage{makeidx}
\usepackage{capt-of}
\usepackage{float}
\usepackage{placeins}
\usepackage{perpage}
\usepackage{mwe}
\usepackage{kantlipsum}
\graphicspath{{images/}}

\usepackage{natbib}
\usepackage{array}
\usepackage{booktabs}
\usepackage{tabu}
\usepackage{dcolumn}
\usepackage{amsmath}
\usepackage{amsfonts}
\usepackage{amssymb}
\usepackage{graphicx}
\usepackage{subfigure}
\usepackage{epsfig}
\usepackage{dcolumn}
\usepackage{bm}

\usepackage{graphicx}
\usepackage{amsfonts}

\def\be{\begin{equation}}
\def\ee{\end{equation}}
\def\bea{\begin{eqnarray}}
\def\eea{\end{eqnarray}}

\newcommand{\pd}{\partial}

\journal{Nuclear Physics B}

\begin{document}

\begin{frontmatter}

\title{Inflation in String Field Theory}

\author[a,b]{Haidar Sheikhahmadi,}
\address[a]{Center for Space Research, North-West University, Mafikeng, South Africa}
\address[b]{School of Astronomy, Institute for Research in
Fundamental Sciences (IPM),  P. O. Box 19395-5531, Tehran, Iran}
\ead{h.sh.ahmadi@gmail.com; h.sheikhahmadi@ipm.ir}
\author[c,d,e]{Mir Faizal,}
\address[c]{Department of Physics and Astronomy, University of Lethbridge, 4401 University Drive, Lethbridge, Alberta T1K 3M4, Canada}
\address[d]{Irving K. Barber School of Arts and Sciences, University of British Columbia - Okanagan, 3333 University Way, Kelowna, British Columbia V1V 1V7, Canada}
\address[e]{Canadian Quantum Research Center, 204-3002, 32 Ave,
Vernon, BC, V1T 2L7, Canada}
\ead{mirfaizal@gmail.com}
\author[f]{Ali Aghamohammadi,}
\address[f]{Department of Physics,Sanandaj Branch, Islamic Azad University, Sanandaj, Iran}
\ead{a.aqamohamadi@gmail.com}
\author[g]{Saheb Soroushfar,}
\address[g]{Faculty of Technology and Mining, Yasouj University, Choram 75761-59836, Iran}
\ead{soroush@yu.ac.ir}
\author[h,i]{Sebastian Bahamonde}
\address[h]{Laboratory of Theoretical Physics, Institute of Physics, University of Tartu, W. Ostwaldi 1, 50411 Tartu, Estonia}
\address[i]{Laboratory for Theoretical Cosmology, Tomsk State University of
Control Systems and Radioelectronics, 634050 Tomsk, Russia (TUSUR)}
\ead{sbahamonde@ut.ee}


\begin{abstract}
In this paper, we analyze the inflationary cosmology using string field theory. This is done by using the zero level contribution from string field theory, which is a non-local tachyonic action.  We will use the non-local  Friedmann equations for this model based on string field theory, and calculate the slow-roll parameters for this model. We will  then explicitly obtain the  scalar and tensorial power spectrum,  their related indices, and the  tensor-to-scalar ratio for this model. Finally, we use cosmological data from Planck 2013 to 2018 to constrain the free parameters in this model, and find that string field theory is compatible with them.
\end{abstract}


\end{frontmatter}



\section{Introduction}
It is possible to study second quantization  of string theory, and this quantum field theory of strings is known as  string field theory.  At a perturbative level this can be done by   finding suitable vertices for string interactions along with string propagators \cite{1a, 2a}. Thus a diagram analogous to the usual Feynman diagrams in quantum field theory can be constructed  for string scattering amplitudes, using string field theory \cite{1, 2}. These Feynman  like diagrams for string field theory can be constructed using second quantized string action, which is constructed using the free string action, along with interaction terms.
However, it is also possible to obtain non-pertubative information about string theory using string field theory. This is because it is possible to calculate the off-shell amplitudes from string field theory using the second quantized action \cite{4a, 4b, 4c, 4d}. This information cannot be obtained   from the standard genus expansion of string scattering amplitudes.  Thus, it is possible to use string field theory to obtain interesting non-trivial information about string theory, which goes beyond string perturbation theory.

It may be noted that even though the string field theory has tachyonic instability, it is possible to analyze  it using tachyonic condensation \cite{a8}.
In tachyon condensation, the system with a tachyonic  instability can  becomes stable near a minimum.
Such condensation in  string field theory has been used to analyze  a system of  coincident brane and  anti-brane pair with  a tachyonic mode, and it has been observed that this system has a configuration with zero energy density and spacetime supersymmetry  \cite{8}.
This is because it  has been demonstrated that  at the  minimum of the tachyonic potential, the  sum of the tension of the brane and the anti-brane cancels the  the negative energy density associated with the tachyonic potential.
The string field theory has been used to study tachyon potential on a D-brane anti-D-brane system in type II string theory. It has been demonstrated that the potential in this case in arbitrary background has a universal form, and it does not depend on the   boundary conformal field theory describing this system \cite{8a}. A classical time dependent solutions for a tachyon on unstable D-branes has been constructed using    string field theory  \cite{8b}. Even though the  action in    string field theory  contains infinite number of time derivatives, it has been possible to obtain  a time dependent solution for this tachyon. This solution is characterized by initial values of this system.
The  string field theory has also been used to unstable branes in the background of a generic linear dilaton \cite{8c}.  It is possible to  obtain a non-pertubative tachyonic vacuum from  the perturbative vacuum  for this system.  This is because the open string tachyon can increase  exponentially along a null direction producing a static state.  Thus, it is possible to dynamically  obtain a  tachyon vacuum in  string field theory from a perturbative vacuum.

 The tachyon condensation has also been  studied in  string field theory action  with a large background field  \cite{8d}.  It has been demonstrated that for string field theory on a background field, the  product  of tachyon field is given by the star product in noncommutative geometry \cite{5d}.
 In fact, it has been demonstrated that noncommutative geometry can be realized in  string field theory using such mechanisms  \cite{ 5b, 5c}.
 In string field theory, there are infinite number of fields, and the action for each of those fields can contain non-local terms. It has been explicitly demonstrated  that the action for zero level of  bosonic string field theory is non-local and contains a tachyonic field \cite{a, b}. The instability of the D-brane in this system are described by this  tachyonic field. The nonlocality in the action is obtained from the  exponential of the d’Alembertian. This nonlocal action has been used to study the spontaneous   symmetry breaking  in such theories \cite{a}.
It has also been demonstrated that such a  infinite derivative non-local theory  can be made ghost free, but   the Green functions for such non-local theories would  modified by an extra term  \cite{d}.  Even though  it can cease a causal effects, those effects  would be  confined in the region of non-locality.

The non-local tachyonic  action  in  string field theory has been used to construct a cosmological solution \cite{c1}.
The equations of motion for this system  are obtained using the Hamilton–Jacobi formalism. Thus,  Friedmann equation are
obtained from this   non-local tachyonic field  in string field theory.  It has also been observed that a general class
of non-local cosmological model can be obtained from string field theory \cite{Arefeva:2005mom, Joukovskaya:2007nq}. In fact, the  Friedmann
equations for such non-local cosmological models has been studied, and it has been demonstrated that these solutions contain  a
 non-singular bouncing solution. As the absence of singularity cannot be a perturbative effect, these non-local
 solutions obtained from string field theory,
 can produce important non-perturbative modifications to the cosmological solutions. It has been observed that non-localities  in a  cosmological solutions can produce an accelerated expansion of the universe
 \cite{c4, c5}. Thus, such non-localities can have important cosmological consequences.
  Now it  would be interesting to investigate inflation in
 string field theory using such non-local cosmological models.
 We would like to emphasize that such non-local  cosmological solution have already been constructed using string field theory  \cite{Arefeva:2005mom, Joukovskaya:2007nq}, and in this paper, we will only use these cosmological  models to analyze  inflation in string field theory .

 {{
 This would be interesting as inflation is able to resolve  the flatness, monopole and horizon problems~\cite{Guth:1980zm,Linde:1981mu,Starobinsky:1980te,Albrecht:1982wi,Yokoyama:1987an,Linde:1983gd,Linde:1985ub,Linde:2005dd}. Remarkably, inflation also gives an explanation for the temperature fluctuations in the cosmic microwave background (CMB), that can be understood as being originated by quantum fluctuations that are stretched out to super-horizon scales and then frozen~\cite{Kolb:1990vq}. Moreover, inflation also explains the large-scale structure formation~\cite{Mukhanov:1981xt}. Inflation is usually understood with a scalar field, called the inflaton, which drives the acceleration.
  This is because in    inflationary cosmology, the inflaton field potential  reaches a local minimum   through supercooling mechanism, and  after a  phase transition, the universe undergoes an  exponential expansion. In inflationary cosmology,    the universe has to have a graceful exit from its inflationary phase to its present state of expansion. Such a graceful exit has been studied using various different inflationary models  \cite{ linde1982new,linde1982coleman,linde1982scalar,albrecht1982cosmology, linde2000inflationary,linde1994hybrid}.
The study of such   inflationary models have led to interesting  various predictions   \cite{revinfl1,revinfl2,revinfl3,revinfl4,revinfl5,revinfl6}. Furthermore,  various  inflationary models with a graceful exist have been constructed using string theory \cite{s12f, s14f, s15f, s16f}.  }}

  {{
In  inflationary cosmology,  inflation ends with  a reheating phase~\cite{Kofman:1997yn,Kofman:1994rk,Bassett:2005xm}. Due to dissipative effects in
warm inflation, it  is also possible for a thermal bath to exist  with the inflaton field
 \cite{w1,w2,w56,w57}. In  such models, physical parameters  can be  randomly distributed. This observation has lead to the development of   distributed mass model in string theory \cite{w3, w4,w5,w6,w7, w8, w48, Ghadiri:2018nok,Sheikhahmadi:2019gzs,Dimopoulos:2019gpz}.
 It may be noted that inflation has been thoroughly studied in string theory \cite{st1, st2, st4, st5, st5a}.
 In fact,  fibre inflationary cosmology has been explicitly constructed in   type IIB  string theory \cite{st10, st20}.  In these models, the
inflaton is a Kahler modulus, and it  controls the size of a  $K3$ divisor fibred a $P^1$ base. String inflation has also been studied using
string gas cosmology \cite{sg12, sg14}. In this model of inflation,  the  string theoretical degrees
of freedom are incorporated in a model of inflationary cosmology  cosmology. This is done using   T-duality
of the string theory.
 As important  models of  string inflationary cosmology have been constructed, it would be interesting to study it in string field theory.
We  also note that inflation with  tachyon potential has been studied in  string theory \cite{c6, c7}. It has been demonstrated that  inflation with such a tachyonic field  can solve the smoothness and flatness problems \cite{c8}.  Inflation in  string theory has also been studied by   compactifying a  bosonic string theory on an internal non-flat space \cite{c9}. In this model, the tachyon potential behaves as the cosmological constant. Thus, tachyons in string theory can also have important application in inflationary cosmology. So, in this paper, we will study inflationary cosmology from  the non-local tachyonic action produced in string field
 theory. }}

The inflationary cosmology has also been studied using non-local Starobinsky gravity
\cite{nonl1}. It has been demonstrated using this non-local model  provides a mechanism for creation and
 thermalization of the matter. This model also fits with the  observational  spectrum of primordial perturbations. In fact, it has been demonstrated that the inflation in these non-local models is driven by both  the gravitons and the scalarons.
 The  magnetogenesis during inflation has been studied by  using a non-local action \cite{El-Menoufi:2015ztk}. It has been demonstrated that this non-local action  incorporates the long distance fluctuations  at the inflationary scale. It has been observed that   the sign of the beta function can have important consequences for this behavior of this system.
 It may be noted that non-local field theories, with   infinitely many
derivatives have also been used to study different aspects of cosmology \cite{xyc1, xyc2, xyc4, xyc5}. These theories resemble the component action obtained from string field theory, as the action obtained from string field theory can be written using a star product. This star product will smear the interactions for component fields  \cite{nl16, nl18}. This will produce non-local interaction terms in the action of those component fields. However, these component fields can be redefined to obtained  actions, with the usual interaction terms and non-local kinetic terms \cite{nl12, nl14}.  Inflation in such theories, with infinite derivatives has also been studied \cite{xyc6, xyc7}. As it is possible to obtain a non-local tachyonic action from string field theory, and  non-local inflationary cosmology can produce important results, we will study inflation using the non-local tachyonic action obtained from string field theory.

This paper is organized as follows: In Sec.~\ref{sect2} we give a brief introduction to non-local string field theory and the corresponding FLRW cosmological equations. 
Then we will study the inflation   in three steps: first, in~~\ref{sec:slow} we show how to derive the slow-roll parameters; then, in~\ref{sec:perturbation} we find out the expressions for the power-spectrum, tensor spectral index and the tensor-to-scalar ratio. Finally, we use these expressions in the last subsection~~\ref{Observations} to constrain the free parameters using recent data from Planck data set. We conclude our main results in Sec.~\ref{sec:conclusions}. Throughout this paper, we use natural units where $c=G=1$ and the metric signature $(-+++)$.

\section{Non-local string field theory and FLRW cosmology}
\label{sect2}
In this section, we will discuss the construction of  a cosmological model based on string field theory. To do that we  will first review the construction of string field theory. This theory is based on  second quantized strings, and it can be constructed using a string field $\ket{\Psi}$.
This   string field can be  express as a sum of the
amplitudes for the string to be  in the various basis states. The zero level  for this string   field  corresponds to  the tachyonic field.  Now in the world sheet string theory, the  Becchi-Rouet-Stora-Tyutin (BRST) charge $Q$ projects out the physical states as
\begin{equation}
Q \ket{\Psi}=0.
\end{equation}
 This can also be written as the following equivalence relation, $\ket{\Psi} \sim \ket{\Psi} + Q \ket{\lambda} $. After second quantization, this equivalence can be   interpreted as a gauge invariance, and the projection of physical states can be interpreted as equations of motion.
Now the free action for free string field would be of the form  $ \bra{\Psi}Q \ket{\Psi} /2$,
where $ \bra{\Psi}$ is the BZP dual of $\ket{\Psi}$.
An interaction term $\langle \Psi,\Psi,\Psi\rangle/3$ can be added to this free action. This term is a  trilinear map which projects  three string fields of total ghost number three to a number. Now motivated by noncommutative geometry, one can define a star product~\cite{b}, satisfying the associativity property, \begin{equation}\Psi_1 * (\Psi_2 * \Psi_3 ) = (\Psi_1* \Psi_2) * \Psi_3\,.
\end{equation}
The BRST charge now acts as a graded derivation on this star product, namely
\begin{equation}Q (\Psi_1 * \Psi_2) = (Q\Psi_1) * \Psi_2 + (-1)^{gn (\Psi_1)} \Psi_1 * (Q \Psi_2)\,,
\end{equation}
where $gn (\Psi_1)$ is the ghost number of $\Psi_1$.  Then,  it is possible to write the action of string field theory using  a star product as follows~\cite{b}
\begin{equation}
S =  \int \frac{1}{2 \alpha^{\prime}}\Psi * Q \Psi + \frac{\mathfrak{g}}{3} \Psi * \Psi * \Psi\,,
\end{equation}
where $\mathfrak{g}$ is the string coupling constant and $\sqrt{\alpha^{\prime}}$ is the string tension. This action is gauge invariant under the following gauge transformations
\begin{equation}
\Psi \to \Psi + Q \Lambda + \Psi * \Lambda - \Lambda * \Psi\,,
\end{equation}
whereas the finite gauge transformations for string field theory can be written as
\begin{equation}
\Psi \to e^{-\Lambda} *(\Psi + Q)* e^\Lambda\,.
\end{equation}
Here the exponential functions  are  defined as
\begin{equation}
e^\Lambda = 1 + \Lambda + \frac{1}{2} \Lambda * \Lambda + \cdots .
\end{equation}
The star product  smears  out the interaction over the string length. Thus, for a particle field in the string field theory spectrum, the interactions will have a
non-local interaction term due to this smearing \cite{nl16, nl18}.
This is because  the  interaction term of the string field theory, which contains a star product,  produces non-local interactions for the component fields.
However, it is possible to use redefined fields, such that the interaction have the usual polynomial form, and the kinetic part of the field picks up this  non-local contribution \cite{nl12, nl14}.
{Thus, it is possible to write the non-local action for the
level zero modes of  string field theory   \cite{ Arefeva:2005mom,Joukovskaya:2007nq}.
It has also been demonstrated that gravitational   singularities can be resolved using string field theory \cite{gr12av}. Additionally, it has  been argued that  string field theory in curved spacetime can be analyzed as a non-linear $\sigma$ model \cite{gr14av, gr15av}.
In fact, cosmological solutions have been obtained using  a non-local action for the  level zero modes of string field theory in curved spacetime  \cite{Arefeva:2005mom,Joukovskaya:2007nq}}
\begin{equation}
\label{full-action}
S=\int \textrm{d}^4x\sqrt{-g}\left[
  \frac{M_{\rm p}^2}{2}R +
  \frac{1}{\mathfrak{g}{\alpha'}^2}\left(
      \frac{{\alpha'}\xi^2}{2}\phi\square\phi+\frac{1}{2}\phi^2 -
      V(e^{\alpha' \kappa_n\square }\phi)\right)
  \right]\,,
\end{equation}
where  $g$ is the determinant of the metric $g_{\mu\nu}$,
 $\square=\nabla_\mu \nabla^\mu=\pd_{\mu}(\sqrt{-g}g^{\mu\nu}\pd_{\nu})/\sqrt{-g}$, $M_{\rm p}$ is the Planck mass,  $\phi$  is the  scalar field,   $\kappa_n$  the  non-local coupling constant,  and  $V$ is the potential. Here the non-locality occurs due to the stat product of string field theory. This non-local action is then coupled to gravity.    To couple this action rigorously to gravity, it would be important to calculate higher level modes of    string field theory. However, as the main motivation in this paper is to analyze the cosmological consequences of the non-locality that occurs in string field theory, we directly couple such a  non-local action to gravity.
 We would  also like to point out that a different   effective action would be produced from  the second quantization of a different string theory. Thus, the specific form of this non-local action depends on the specific string field theory \cite{Arefeva:2005mom,Joukovskaya:2007nq}.
 It is then possible to rewrite this action in dimensionless variables as  \cite{Arefeva:2005mom,Joukovskaya:2007nq},
\begin{equation}
\label{action}
S=\int \textrm{d}^4x\sqrt{-g}\left(\frac{\tilde{M}_{\rm p}^2}{2}R+
\frac{\xi^2}{2}\phi\square\phi+\frac{1}{2}\phi^2-V(\Phi) \right)\,,
\end{equation}
where
$\phi$ is a dimensionless scalar field, which is used to define an effective scalar field as $\Phi=e^{\kappa \square}\phi$, $\tilde{M}_{\rm p}^2=\mathfrak{g}{\alpha'}{M_{\rm p}^2}  $ and $\kappa={\alpha'} \kappa_n  $.
The Friedmann-Lema\^itre-Robertson-Walker (FLRW) metric in Cartesian coordinates
\begin{equation}
\label{mFr}
d\textrm{s}^2=-\textrm{d}t^2+a^2(t)\left(\textrm{d}x^2+\textrm{d}y^2+\textrm{d}z^2\right)\,,
\end{equation} 
where $t$ is the cosmological time and $a(t)$ the scale factor. As we are considering  a spatially homogeneous configuration,  the
Beltrami-Laplace operator takes the form
$ \square=-\partial_t^2-3H(t)\partial_t$, where $H(t)=\dot{a}(t)/a(t)$ is the Hubble parameter. Note that dots mean diferentiation with respect to the cosmic time $t$. For simplicity, we will also use the  notation $\mathcal{D}^2=-\square=\partial_t^2+3H(t)\partial_t$. The extra term $3H(t) \partial_t$ appears due to the  curved spacetime and the choice of the Levi-Civita connection acting on the covariant derivative, as in Genereal Relativity.

Varying the action~\eqref{action} with respect to the scalar field $\phi$ (using  the definition of the effective scalar field $\Phi$), we obtain the following equation of motion
\begin{equation}
\label{eom-phi-H}
(1-\xi^2\mathcal{D}^2)e^{2\kappa \mathcal{D}^2}\Phi = V^{\prime}(\Phi)\,,
\end{equation}
here  $d\Phi/d\phi=e^{-\kappa\mathcal{D}^2}$ has been used and primes denote differentiation with respect to the effective scalar field $\Phi$. The above equation is a modified Klein-Gordon equation. {{ Even though  during the   level truncation procedure, the value of $\xi$ used is  $
\xi^{2}=-1/{4 \log ({4}/{3 \sqrt{3}})}$, it has been argued that various different values of $\xi$ can produce various different physical models  \cite{Joukovskaya:2007nq}. In fact, it has been observed that  $\xi =0$ corresponds to  $p=3$ for  $p$-adic strings \cite{Joukovskaya:2007nq, xyc5}.
However, the $p$-adic strings have not been second quantized, and so such values cannot be directly obtained from string field theory.
As it is possible to consider different values of $\xi$,    we will find a  bound for the  value of $\xi$ from observational data.   }}



The Friedmann equations can be obtained by  varying the action~\eqref{action} with respect to the metric and then replace the FLRW metric~\eqref{mFr}, yielding~\cite{Arefeva:2005mom,Joukovskaya:2007nq,Barnaby:2007yb}
\begin{eqnarray}\label{Fried1}
3H^2=\frac{1}{\tilde{M}_{\rm p}^2}~{\cal {E}}_{\rm tot}~, &&
2\dot H+3H^2={}-\frac{1}{\tilde{M}_{\rm p}^2}~{\cal P}_{\rm tot}~,
\end{eqnarray}
where the subscript $\rm ``tot"$ indicates the sum of both local and  non-local contributions to the energy density and pressure.  Therefore, due to the star product of string field theory, there is an additional  non-local contribution to the energy density and the  pressure, namely~\cite{Joukovskaya:2007nq,Barnaby:2007yb}
\begin{eqnarray}\label{E-tot}
\mathcal{E}_{\rm tot}=
\mathcal{E}_k+\mathcal{E}_p
+~\tilde{{\cal E}}_1~+~\tilde{{\cal E}}_2\,, \label{P-tot}
\quad \mathcal{P}_{\rm tot}=
\mathcal{E}_k-\mathcal{E}_p
-~\tilde{{\cal E}}_1~+~\tilde{{\cal E}}_2\,,
\end{eqnarray}
where $
 {\cal E}_k=\xi^2(\partial_t\phi)^2 /2, $ $ {\cal E}_p={}-\phi^2+V(\Phi)/2,$ and
\begin{subequations}
\begin{eqnarray}\label{E1-nonloc}
\tilde{{\cal E}}_1&=&\kappa\int_{0}^{1} \textrm{d} \varrho \left(e^{-\kappa\varrho \mathcal{D}^2}V^{\prime}(\Phi)\right)
 \left(\mathcal{D}^2 e^{k\varrho\mathcal{D}^2}\Phi\right)\,,
\label{E2-nonloc}\\
\tilde{{\cal E}}_2&=&-\kappa\int_{0}^{1}  \textrm{d} \varrho\left(\partial_t e^{-\kappa \varrho\mathcal{D}^2}
V^{\prime}(\Phi)\right) \left(\partial_t e^{\kappa \varrho \mathcal{D}^2}\Phi\right)~.
\end{eqnarray}
\end{subequations}
Now  substituting Eq.~(\ref{eom-phi-H}) into the above equations, we obtain
\begin{subequations}
\begin{eqnarray}\label{E1-nonloc-Operator}
\tilde{{\cal E}}_1&=&\kappa\int_{0}^{1} \textrm{d} \varrho \left( (-\xi^2\mathcal{D}^2+1)e^{(2-\varrho)\kappa\mathcal{D}^2}\Phi\right)
 \left(\mathcal{D}^2 e^{\kappa\varrho\mathcal{D}^2}\Phi \right)~,
\\
\label{E2-nonloc-Operator}
\tilde{{\cal E}}_2&=&-\kappa\int_{0}^{1}  \textrm{d} \varrho\left(\partial_t  (-\xi^2\mathcal{D}^2+1)e^{(2-\varrho)\kappa \mathcal{D}^2}\Phi
\right) \left(\partial_t e^{\kappa \varrho \mathcal{D}^2}\Phi\right)~.
\end{eqnarray}
\end{subequations}
Using  Eq.~(\ref{Fried1}), it is possible to  obtain   the time evolution of the Hubble parameter, which is
\begin{equation}
\label{Fried2}
\dot H=-\frac{1}{2\tilde{M}_{\rm p}^2}~({\cal P}_{\rm tot}+{\cal E}_{\rm tot})~.
\end{equation}
Then, by substituting Eqs.~\eqref{E1-nonloc-Operator} and \eqref{E2-nonloc-Operator} into Eq.~\eqref{E-tot} and then by integrating the second Friedmann equation~\eqref{Fried2},  it is possible to get
\begin{equation}
\label{fr2-proint}
H=-\frac{1}{\tilde{M}_{\rm p}^2} \int_0^t \textrm{d} {t^\prime} ~\left[\frac{\xi^2}{2}(\partial_{t^\prime}\phi)^2
-\kappa\int_{0}^{1} \textrm{d} \varrho \left(\partial_ {t^\prime}(-\xi^2 \mathcal{D}^2+1) e^{(2-\varrho)\kappa\mathcal{D}^2}
\Phi\right) \left(\partial_{t^\prime} e^{\kappa \varrho \mathcal{D}^2}\Phi\right)\right]~.
\end{equation}

 To solve this equation,  we need to deal with the non-local operator
$e^{k \varrho\mathcal{D}^2}$. To do this, we can follow the same approach done in~\cite{Joukovskaya:2007nq}, where the authors assumed an ansatz for the effective scalar field, and use it to obtain  an exact solution for the Hubble parameter. After finding this exact solution, one can then use the Klein-Gordon equation~\eqref{eom-phi-H}, to find out the form of the potential.

Before going further more, let us briefly discuss the properties and conditions that our definitions should obey. By acting the exponential operator, i.e. $e^{\kappa\varrho\mathcal{D}^2}$, on a specific arbitrary function $F(t)$, one finds a solution which looks like a diffusive partial differential equation with
corresponding initial and boundary conditions. According to~\cite{Joukovskaya:2007nq}, we then require the following
conditions,
\begin{eqnarray}
\label{f-rho-t}
\frac{\pd}{\pd \varrho} F(\varrho, t)  = &{\mathcal{D}^2} F(\varrho, t)~, \,\,\,\,\,\,\,
F(0, t)  =  F(t)~, \,\,\,\,\,\,\,
F(\varrho, \pm \infty) =  F(\pm \infty)~,\label{f-rho-boundary}
\end{eqnarray}
where $-\infty < t < +\infty$ and $\varrho \geqslant 0$. One can obtain that the exponential operator acts as $e^{\kappa \varrho\mathcal{D}^2}F(t)=F(\varrho, t)$,  and then, such solution exists and is unique for wide classes of the Hubble parameter and
$F$~\cite{Barnaby:2007yb}. We can then follow the same idea done in~\cite{Joukovskaya:2007nq} by assuming the following ansatz which satisfies the above conditions:
\begin{equation}\label{ansatz}
    F(\varrho, t)=\Phi(\varrho, t)=\alpha^n e^{\beta\varrho} t^{nq}~,
\end{equation}
where $\alpha,~\beta,~q$ and $n$ are the parameters of the model that will be fitted afterwards using data from observations. {It may be noted that we have used a specific form of this ansatz involving  $\alpha,~\beta,~q$ and $n$. This specific form was motivated by the non-local action  \cite{Arefeva:2005mom,Joukovskaya:2007nq,Barnaby:2007yb}, but the values of these parameters are not fixed by string field theory.
So,  will be find bounds on their values from observational data. }
Now considering this ansatz, one can easily solve Eq.~(\ref{fr2-proint}) for the Hubble parameter, yielding
 \begin{eqnarray}\label{Hubble-exact}
H(t)=\frac{\left( n q \alpha^{n}e^{ \beta }\right)^2 \left(2 \kappa \left( \xi ^2-\beta\right)+\xi ^2\right) }{2 \tilde{M}_{\rm p}^2 (1-2 n q)}t^{2 n q-1}~.
\end{eqnarray}
Let us here remark that we will further assume that $n\neq 1/(2q)$ since this specific case is another branch of solution which gives a different Hubble parameter. Now, we can replace this expression into the Eq.~\eqref{eom-phi-H}, and obtain the potential,
 \begin{eqnarray}\label{Potential-Exact}\nonumber
  V(t)&=&\frac{9{{e}^{2\beta }}{{\alpha }^{3n}}{{({{n}^{2}}{{q}^{2}}{{\alpha }^{n}}{{e}^{\beta }}\xi )}^{2}}\left( 2\kappa ({{\xi }^{2}}-\beta )+{{\xi }^{2}} \right)}{3\tilde{M}_{\text{p}}^{2}(5nq-2)}{{t}^{5nq-2}} \\
 & +&\frac{{{e}^{2\beta }}{{\alpha }^{3n}}(2nq-1)}{3}(\frac{6{{n}^{2}}{{\xi }^{2}}{{q}^{2}}(1-2nq)}{3nq-2}{{t}^{-2+3nq}}+{{t}^{3nq}}).
\end{eqnarray}
Here again we further assumed that $n\neq 2/(3q)\neq 2/(5q) $ since they are different branches for the model. Now Eqs.~\eqref{Hubble-exact} and \eqref{Potential-Exact} are exact solutions of our model for the ansatz assumed. In the next section we are going to investigate the slow-roll conditions to obtain the time for the horizon exit $t_{\rm e}$ and consequently $\Phi$ at the horizon exit. In fact, since the first slow-roll parameters are defined based on the acceleration, which contains time derivatives of the Hubble parameter, we expect that in the end of the inflationary phase, they should be equal to the unity. Consequently, by virtue of the e-folding definition, the time at the horizon exit can be calculated as well.
\section{Slow-roll non-local string field inflation}\label{sec:slow}
Now we can use the non-local solution obtain from string
field theory to analyze the
observational constraints on inflation in string field theory.
In order to investigate the observational constraints from the inflationary era, we introduce the first, and second slow-roll parameters, which are defined as
\begin{eqnarray}
\varepsilon _1 &:=&  - \frac{\dot {H}}{H^2}=\Big(\frac{ \tilde{M}_{\rm p}(1-2 n q)}{n q \alpha^{n}e^{\beta } }\Big)^2\frac{2}{\left(2 \kappa \left(\xi ^2-\beta \right)+\xi ^2\right)}\,t^{-2nq}~,\label{slowroll1}\\
\varepsilon_2&:=&\frac{\dot{\varepsilon _1}}{H{\varepsilon _1}}
\equiv  \frac{\ddot{H}}{H\dot{H}}-\frac{2\dot{H}}{H^2}=\frac{4(2 n q-1)\tilde{M}_{\rm p}^2}{n q (\alpha^{n}e^{\beta })^2  \left(2 \kappa \left( \xi ^2-\beta\right)+\xi ^2\right)}\,t^{-2nq}~.\label{slowroll2}
\end{eqnarray}
As mentioned before, the time of the end of inflation $t_{\rm e}$ can be obtained by equating $\varepsilon_1$ to one, giving
\be\label{Time-End}
t_{\rm e}= \left(\frac{\sqrt{2}\tilde{M}_{\rm p} (2 n q-1)}{\alpha ^{n}e^{\beta }n q  \sqrt{2 \kappa \left(\xi ^2-\beta\right)+\xi ^2}}\right)^{1/(n q)}~.
\ee
Now from the ansatz~\eqref{ansatz}, one notices that the effective scalar field can be written as $t(\Phi)=(\Phi /\alpha)^{1/(n q)}$, and then by using the definition introduced in \eqref{slowroll1}, it follows that in terms of the effective field potential, the first slow-roll parameter can be expressed as
\begin{eqnarray}\label{vsrp}
\varepsilon_1& =& {\tilde{M}_{\rm p}^2 \over 2} {V^{\prime 2}(\phi) \over V^2(\phi)}~.
\end{eqnarray}
In calculations related to the power spectrum and spectral indices, it is also useful to introduce  another slow-roll parameter defined in terms of the potential as:
\begin{eqnarray}\label{etabeta}\nonumber
\eta &:=& {\tilde{M}_{\rm p}^2} \; {V''(\Phi) \over V(\Phi)}= \frac{6M_{\rm p}^2(2n q-1)}{\Phi ^2} \\
&\times&\frac{ \left(3 (e^{\beta } n \xi  q\alpha ^{ n})^2 \left[2 k (\xi ^2-\beta)+\xi ^2\right] \left(\frac{ \Phi }{\alpha }\right)^{2 }\right)-2\tilde{M}_{\rm p}^2 \left[ (n q-1)(2n q-1)\xi^2 +\left(\frac{ \Phi }{\alpha }\right)^{\frac{2 }{n q}}\right]}{\frac{9 e^{2 \beta }  (e^{\beta } n^2  q^2\xi \alpha ^{ n})^2  \left[2 k (\xi ^2-\beta)+\xi ^2\right] }{5 n q-2}\left(\frac{ \Phi }{\alpha }\right)^{2}-\tilde{M}_{\rm p}^2 \left[\frac{ (1-2n q)^2 (n  q \xi)^2 }{3n q-2}+(1-2n q)\left(\frac{ \Phi }{\alpha }\right)^{\frac{2 }{n q}}\right] }\,.
\end{eqnarray}
The amount of cosmic expansion during  inflation is measured through the number of e-folding $\mathcal{N}$, defined as
\bea\label{efoldtau}
\hspace{-0.3cm}\mathcal{N }&=& \int_{t_\star}^{t_{\rm e}} H(t) \; \textrm{d}t =\frac{n q (e^{\beta } \alpha^{ n})^2  \left[\xi ^2+2 \kappa \left(\xi ^2-\beta\right)+\xi ^2\right] }{4\tilde{M}_{\rm p}^2(1-2 n q)}\, \left({t_{\star}}^{2 n q}-{t_{\rm e}}^{2 n q}\right)~,
\eea
where $t_{\star}$ is the time at the horizon crossing, and we have used \eqref{Time-End} to introduce the time of the end of inflation $t_{\rm e}$. One can further invert the above equation to obtain $t_{\star}$ in terms of the e-folding numbers, which yields in
\bea\label{Time-exit}
t_\star=\alpha^{-1/q}\left(\frac{4 e^{-2 \beta } \tilde{M}_{\rm p}^2 \mathcal{N}}{n q \left(2\kappa( \beta  - \xi ^2)-\xi ^2\right)}-\frac{8 e^{-2 \beta } \tilde{M}_{\rm p}^2 \mathcal{N}}{2\kappa( \beta  -\xi ^2)-\xi ^2}+\left(\alpha t_{e}^q\right)^{2 n}\right)^{1/(2 n q)}~ .
\eea
To avoid having a cumbersome expression, we have just wrote $t_{\star}$ in terms of $\mathcal{N}$ and the time of the end of inflation $t_{\rm e}$, however, when one wants to use this expression to constrain the model using data, one would need to replace the value of $t_{\rm e}$ expressed in Eq.~\eqref{Time-End}.

\section{Scalar Perturbations and power-spectrum in non-local string field inflation}\label{sec:perturbation}
In this section, we will derive the most important observable quantities in inflation, the power-spectrum, tensorial spectral index and the tensor-to-scalar ratio. Since our theory can be rewritten as a particular scalar tensor theory in an Einstein frame, the computation is standard. For more details about this, see \cite{Mukhanov:2005sc}. First, we start by perturbing the scalar field around a homogeneous background,
\be\label{decompo-Phi}
\Phi(t, \mathbf{x})=\Phi_0(t)+\delta\Phi(t,\mathbf{x})~.
\ee
Then, we need to take perturbations for the metric. In the longitudinal gauge, the perturbed metric is usually written as
\begin{equation}
\textrm{d}s^2 = -\big(1 + 2A(t,\mathbf{x}) \big) \textrm{d}t^2 + a^2(t) \big(1 - 2\Psi(t,\mathbf{x})\big)\delta_{ij} \textrm{d}x^i \textrm{d}x^j~,
\end{equation}
where we denoted the two scalar perturbations by $A(t,\mathbf{x})$ and $\Psi(t,\mathbf{x})$.  For a diagonal energy-momentum tensor (no fluxes, or heat), $\delta T^i_j \propto \delta^i_j$, one can conclude that $A(t,\mathbf{x})=\Psi(t,\mathbf{x})$. It is important to note that this equation also holds for this string field theory since we only have that the scalar field is minimally coupled to gravity, i.e., our theory can be understood as being working in an Einstein frame. In other theories, one can get a  gravitational slip equation where $A(t,\mathbf{x})\neq \Psi(t,\mathbf{x})$. For more discussion about this, see~\cite{Clifton:2011jh}. \\

By perturbing Eq.~(\ref{eom-phi-H}) and then using~\eqref{decompo-Phi} in the momentum space and conformal time $\tau=\int dt/a(t)$, one gets the following  Mukhanov equation,
\be\label{pert-Muck-Sasa}
(\partial _\tau ^2 + {\left| \mathbf{k} \right|^2}){e^{2\kappa(\partial _\tau ^2 + 2\frac{{{d_\tau }a}}{a} + {{\left| \mathbf{k} \right|}^2})}}{v_k} - \frac{{\partial _\tau ^2z}}{\xi^{2}z}{v_k} = 0~,
\ee
where ${v_k} = a\delta \Phi ,$ and $\partial _\tau ^2z/z=\xi^{2}(aH)^2\left(2-3\eta+2\varepsilon_1\right)$ is the well-known Mukhanov-Sasaki variable, or minus mass square  factor. The solution of this equation is equivalent to the quantization of a free scalar field, ${v_k}$ with a time dependent mass in a flat spacetime (Minkowski)~\cite{Mukhanov:2005sc}.
To obtain the above equation we have used $d_\tau ^2a/a = {(aH)^2}(2 - \varepsilon_1 )$ and $\partial _\tau ^2V \approx 3{H^2}(\eta  + \varepsilon_1 ) - 1$. One should note that to find the final expression in the Mukhanov-Sasaki variables, besides these definitions, we needed to take into account the background perturbation ${\xi^{2}(aH)^2}(6 \varepsilon_1)$.

 Using the quantum field  $\delta \Phi$, one can write down the Fourier transformation of the inflaton fluctuation as
\be\label{inflaton-fluctuation-Fourier}
\delta \Phi (t,\mathbf x) = \int {\frac{{{\textrm{d}^3}k}}{{{{(\sqrt {2\pi } )}^3}}}} {e^{i\mathbf k\cdot \mathbf x}}{\sigma _k}(t)~,
\ee
where
\be\label{Sigma-Fourier}
{\sigma _{\mathbf k}}(t) = {a_{\mathbf k}}\delta {\Phi _{\mathbf k}}(t) + a_{ - {\mathbf k}}^\dag \delta \Phi _{ - {\mathbf k}}^ * (t)~,
\ee
and the mode functions $\delta {\Phi _{\mathbf k}}(t)$  are given by
\be\label{Mode-Function}
v_k =a\delta \Phi= {1 \over 2}\; e^{i{\pi \over 2}(\nu + {1 \over 2})} \sqrt{\pi \over {a H}} \; H_\nu^{(1)}\Big(\frac{k}{a H}\Big)~.
\ee
These mode functions usually can be determined by using its standard relationship with curvature perturbation, say $\zeta$, namely $v_k \equiv z \zeta$, and $\zeta$ is the curvature perturbation given by $\zeta = A+ H \; \delta\Phi/ \dot{\Phi}_{0}.$ Here, based on the Mukhanov variable definitions and \eqref{Mode-Function},  the order of the Hankel functions becomes
\be\label{Hankel-orde}
\nu^2-\frac{1}{4}\approx\xi^{2}(2-3\eta+2\varepsilon_1)~.
\ee
The spectrum of curvature perturbation is then defined as
\begin{equation}
  \mathcal{P}_{\rm s} = {k^3 \over 2\pi^2}\; \left| \zeta \right|^2 = {k^3 \over 2\pi^2}\; \left| {v_k \over z} \right|^2\,.
\end{equation}
On superhorizon scales $|k \tau|\ll 1$, the Henkel function asymptotically behaves as
\begin{equation}
  \lim_{-k\tau \rightarrow \infty}H_\nu^{(1,2)}(x) = \sqrt{{2 \over \pi}} \; {1 \over \sqrt{2 k}} e^{\mp (i k\tau + \delta)}\,,
         \qquad  \delta={1 \over 2} \Big( \nu + {1 \over 2} \Big)\,.
\end{equation}
Then, the spectrum of curvature perturbation on superhorizon scales will be
\begin{equation}\label{psspectrum}
  \mathcal{P}_{\rm s} =  \mathcal{P}_{0\rm s}\Gamma_0 \times\; \left( H \over 2\pi  \right)^2 \;  \left( k \over a H \right)^{3-2\nu}~,
\end{equation}
where $\mathcal{P}_{0\rm s}$ is a constant that should be determine using observations and \begin{equation}
    \Gamma_0=\left( 2^{\nu-{3 \over 2}} \Gamma(\nu) \over \Gamma(3/2) \right)^2~.
\end{equation}  Now from Eq.~(\ref{psspectrum}) one can immediately read the scalar spectral index,
\begin{equation}\label{sindex}
  n_{\rm s} -1= {3-2\nu}~.
\end{equation}
We can then define the tensor power spectrum $ \mathcal{P}_{\rm t}$ as
\begin{equation}\label{ptspectrum}
  \mathcal{P}_{\rm t}: =  \frac{2}{\tilde{M}_{\rm p}^2}\left( H \over \pi  \right)^2 \;,
\end{equation}
and then, the tensor spectral index, $ n_t$
\begin{equation}\label{tindex}
  n_{\rm t} := \frac{{d\ln{\mathcal{P}{\rm t}}}}{d\ln{k}}~.
\end{equation}
Finally, the  tensor-to-scalar ratio will be the quotient between the power spectrums,
\begin{equation}\label{tsratio}
 r:=\frac{\mathcal{P}_{\rm t}}{\mathcal{P}_{\rm s}} \;.
\end{equation}
This quantity is a a direct measure of the energy scale of inflation (around GUT energy scales), and based on  cosmological observations by Planck-2018, it is known an upper bound on it as, $r < 0.064$, and there is still some uncertainty on its the direct measurement~\cite{pl2018a,pl2018b}.

\section{Constraints on the model with Planck data sets and Horizon exit investigations}
\label{Observations}
We will now use the results found in the previous sections for our proposed non-local string field theory model, to observationally constrain the parameters. To do this, we will use observational data from Planck 2013, 2015 and 2018 satellites.  These data  sets have been widely used to constrain or even rule out possible theoretical predictions of different inflationary models. The most important set of inflationary parameters are then displayed in Table~\ref{Tab1}. One of the most remarkable predictions of the standard inflationary models is to predict  the non Gaussianity of the power-spectrum and quantum perturbations due to initial bangs~\cite{  Lyth:1998xn,Guth:2000ka,Chen:2009we,Chen:2017ryl,Golovnev:2008cf,Adshead:2012kp,weinberg,PhysRevLett112011302,Emami:2013lma,Sheikhahmadi:2019xkx}. Amongst these fluctuations, the scalar perturbations are considered as the cornerstone and the primary seeds leading to large structure formation \cite{Lidsey:1995np,Bassett:2005xm}. In addition to these, obviously, the tensor perturbations can be interpreted as a source of the gravitational waves~\cite{Grishchuk:1974ny,Starobinsky:1979ty,Allen:1987bk,Sahni:1990tx,Souradeep:1992sm,Giovannini:1999qj,Sami:2004xk,Hossain:2014zma,Geng:2015fla,Hossain:2014coa,Cai:2014uka}. During the inflationary epoch, these perturbations can stretch out the horizon and remain invariant due to its scale invariant property. The observational data has almost shown an  scale invariant spectrum for the curvature perturbations~\cite{pl2018a}. This scale invariant feature is described from the scalar spectral index $n_{\rm s}$ appearing in the exponent of the spectrum of curvature perturbations, that can be written using~\eqref{psspectrum} and \eqref{sindex} as
\begin{equation}
    \mathcal{P}_{\rm s} = \mathcal{A}_{\rm s}^2 \left(k \over a H \right)^{n_{\rm s}-1}~,
\end{equation} where $\mathcal{A}_{\rm s}^2$ is the amplitude of the scalar perturbations at the horizon exit, $t_{\rm e}$, say $k = aH$~\cite{pl2018a,pl2018b}. For the choice of $n_{\rm s}=1$, the amplitude of scalar perturbation is exactly scale invariant, however, the latest observational data coming from combining Planck with the WMAP polarization data, implies that $\ln\left( \mathcal{A}_{\rm s} \times 10^{10} \right) \approx3.044 \pm 0.014 $ and $n_{\rm s}=0.9649 \pm 0.0042$. This means that observations have taught us that there is not a scale invariant perturbation~\cite{pl2018a,pl2018b}. To measure the tensor perturbations, the usual indirect procedure  is to considering the parameter  tensor-to-scalar ratio defined in~\eqref{tsratio}.

\begin{figure}[H]
\centering
\includegraphics[scale=.4]{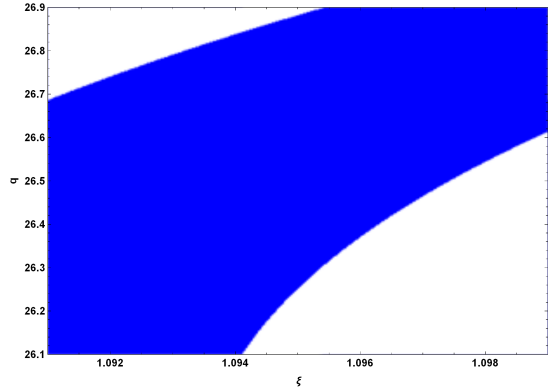}
\caption{{\it{
 This plot shows a set of $(q,\xi)$ pair that can give the best fitted values based on the  $(r-n_{\rm s})$ diagram originated from Planck $2018$ observational data. Here the dark blue color shows the results for $68\%$ CLs of Planck $2015$ data.. To plot this diagram we considered the results of Table I and $\tilde{M}_{\rm p}=1$.}}}
\label{StrongPlotNsraa}
\end{figure}

\begin{figure}[ht]
\centering
\includegraphics[scale=.61]{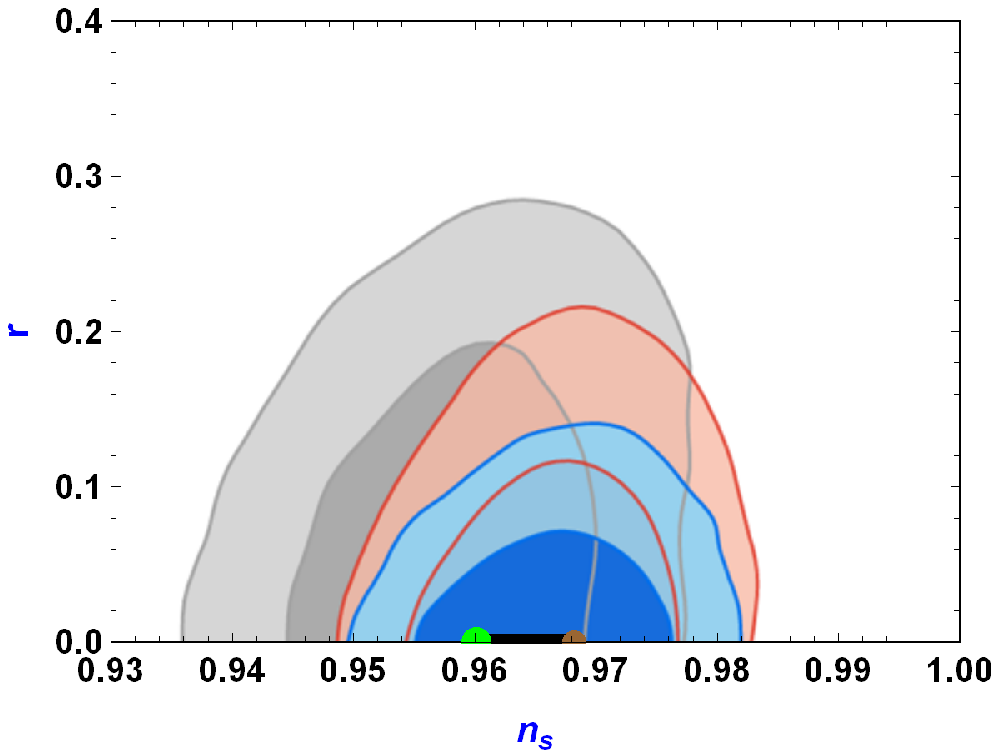}
\includegraphics[scale=.61]{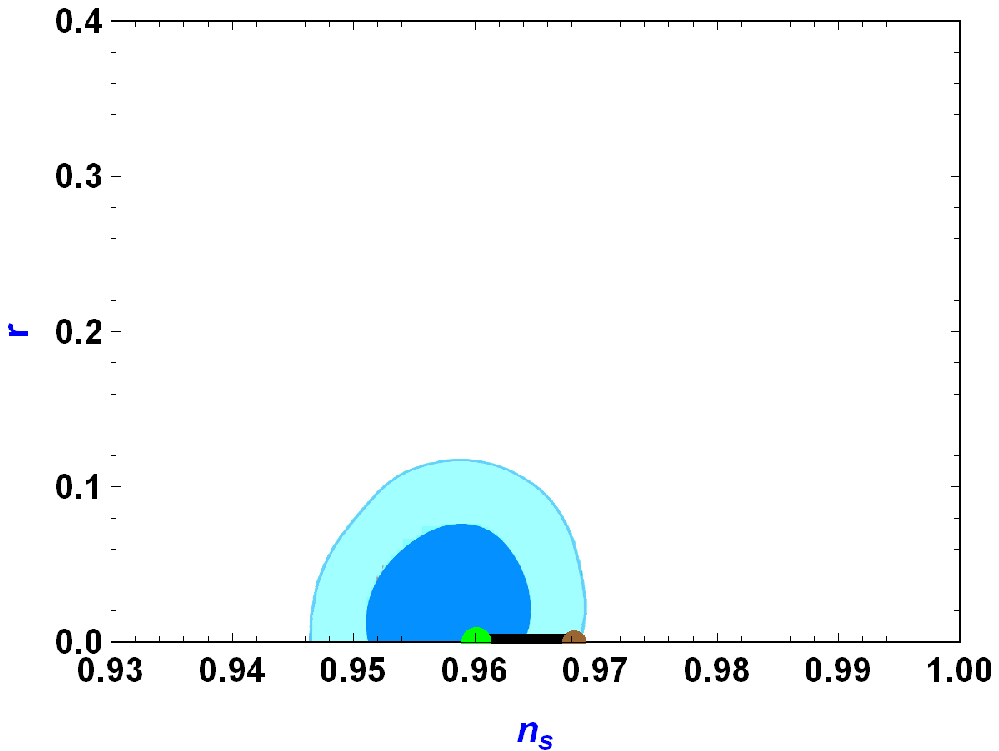}
\caption{{\it{
 The $r-n_s$ diagram shows  Prediction of
the theoretical results  in non-local string field for free parameters of Table 1, $M_{\rm p}=1$ and $\mathcal{N}=65$,
in comparison to the observational data risen by Planck $2013$, $2015$ and 2018 data sets. In the left figure, the likelihood  of  Planck 2013 are indicated with grey contours, Planck TT+lowP with red contours, and Planck TT,TE,EE+lowP (2015) with blue contours. And in the right panel, the results of Planck 2018 are indicated by dark and light blue colours referring $68\%$ and $95\%$ confidence levels respectively. In both figures the thick black lines
refers the predictions of theoretical results in which small and large circles are the values of $n_{\rm s}$ at the number of e-folds $\mathcal{N}=60,~\mathcal{N}=65$ respectively.
}}}
\label{fignsrstrong}
\end{figure}

\begin{figure}[H]
\centering
\includegraphics[scale=.65]{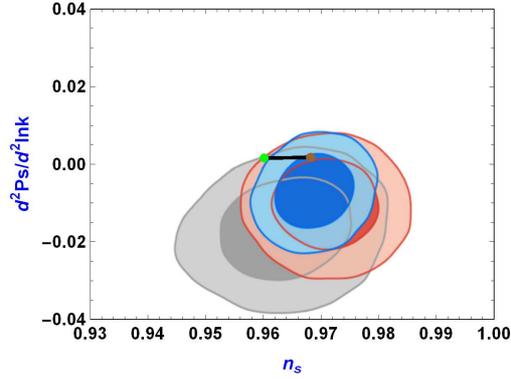}
\caption{{\it{
 This is the $d{n_{\rm s}}/dln k - {n_{\rm s}}$ diagram which indicates the running of the parameter $n_{\rm s}$. In this figure the thick black line
depicts the predictions of our model in which small and large circles are the value of $n_{\rm s}$ at the number of e-fold $\mathcal{N}=60,~ and ~\mathcal{N}=65$ respectively. To plot this diagram we used the free parameters of table I, $M_{\rm p}=1$, $\mathcal{N}=65$.
In this diagram  the grey contours indicated for likelihood  of  Planck 2013, Planck TT+lowP showed with red contours and the blue contours considered for Planck TT,TE,EE+lowP (2015).}}}
\label{strongfignsdns}
\end{figure}

Now we will examine the if our model can predict inflation according to current cosmological data. To do this, we will need to find suitable free parameters of the model and then constrain them accordingly. In doing so, we first need to rewrite all our equations based on the horizon exit time $t_\star$, that in our model is given in Eq.~\eqref{Time-exit}. This equation can be explicitly written only in terms of the e-folding if one replaces Eq.~(\ref{Time-End}) into Eq.~(\ref{Time-exit}), yielding
\bea\label{Time-exit-00}
t_\star(\mathcal{N})= \alpha^{-1/q}  \left(\frac{2e^{-2 \beta } M_{\rm p}^2 (2 n q-1) \left(2 \alpha  n {\mathcal{N}} q+\alpha ^{2 n} (2 n q-1)\right)}{\alpha  n^2 q^2 \left(2 k \left(\xi ^2-\beta \right)+\xi ^2\right)}\right)^{1/(2 q n)}~ .
\eea
Let us  start rewriting the power spectrum equation~\eqref{psspectrum} at the horizon exit, which becomes
\begin{equation}\label{ps0spectrum-exit}
  \mathcal{P}_{0\rm s} = \mathcal{P}_{\rm s}^\star \left( 2^{\nu^\star-\frac{3}{2}}\, \Gamma(\nu^\star) \over \Gamma(3/2) \right)^{-2} \; \left( H^\star \over 2\pi  \right)^{-2}~.
\end{equation}
Note here that we use the superscript $\star$ to  indicate the horizon exit. Hereafter, we will consider that the observed value of the power spectrum is $\mathcal{P}_{\rm s}^\star=2.17*10^{-9}$, and $ \mathcal{P}_{0\rm s}$ can be estimated by using data.
Now, using~\eqref{Time-exit-00}, one finds that the slow-roll parameters~\eqref{slowroll2} and \eqref{etabeta} become
\begin{eqnarray}
\label{epsiolonexit}
 \varepsilon _1^\star(\mathcal{N}) &=&\Big(\frac{ M_{\rm p}(1-2 n q)}{n q\alpha^{n}e^{\beta } }\Big)^2\,\frac{2}{\left(2 \kappa \left( \xi ^2-\beta \right)+\xi ^2\right)} \,t_\star^{-2nq}~,\\
 \label{etaexit}
 \eta^\star(\mathcal{N})&=&\frac{6\tilde{M}_{\rm p}^2 (2 n q-1)\left(3 (e^{ \beta } n \xi  q)^2 \left(2 k \left(\xi ^2-\beta \right)+\xi ^2\right) \left(\alpha  t_\star^q\right)^{2 n}+\tilde{M}_{\rm p}^2 \left(t_\star^2-2 \xi ^2 (n q-1) (2 n q-1)\right)\right)}{\left(\alpha  t_\star^q\right)^{2 n} \left(\frac{9 (e^{ \beta } n^2 \xi q^2)^2 \left(2 k \left(\xi ^2-\beta \right)+\xi ^2\right) }{5 n q-2}\left(\alpha  t_\star^q\right)^{2 n}+\tilde{M}_{\rm p}^2 (2 n q-1) \left(\frac{6 (n \xi  q)^{2} (1-2 n q)}{3 n q-2}+t_\star^2\right)\right)}~.
\end{eqnarray}
Therefore, using Eqs.(\ref{Hankel-orde}),~\eqref{sindex}, \eqref{epsiolonexit} and \eqref{etaexit}), one can easily  find the power spectral index at the horizon crossing time, which becomes
\begin{equation}\label{sindexexit}
  n_{\rm s}^\star -1= {3-2\left(\frac{1}{4}+\xi^{2}(2-3\eta^\star+2\varepsilon_1^\star)\right)^{1/2}} ~.
\end{equation}
Now for the tensorial parameters, one can use Eqs.~\eqref{ptspectrum} and \eqref{tindex}, to get the tensorial power spectrum and the spectral index,
\begin{eqnarray}
\label{ptspectrum}
  \mathcal{P}_{\rm t}^\star &=&  \frac{2}{\tilde{M}_{\rm p}^2}\left( H^\star \over \pi  \right)^2 \;,\\
  \label{tindex}
  n_{\rm t}^\star &=& \frac{{d\ln{\mathcal{P}_{\rm t}}}}{d\ln{k}}\Big|_{t_\star}=-\Big(\frac{ \tilde{M}_{\rm p} (1-2 n q) }{n q \alpha^{n}e^{\beta }}\Big)^{2}\,\frac{4}{ \left(2 k \left(\xi ^2-\beta \right)+\xi ^2\right)}\, \ t_\star^{-2nq}~.
\end{eqnarray}
For the definition of the tensor-to-scalar ratio (see Eq.~\eqref{tsratio}), we also trivially find that,
\begin{equation}\label{tsratio2}
 r^\star=\frac{\mathcal{P}_{\rm t}^\star}{\mathcal{P}_{\rm s}^\star}\;.
\end{equation}
Now, we have all the ingredients to analyze if our model can predict inflation or not. In Fig.~\ref{StrongPlotNsraa}, we constrain the parameters $q$ and $\xi$ which comes from our model. This is done using Planck 2018 observational data, which is summarized in Table~\ref{Tab1}. The $r-n_{\rm s}$ diagram for our model is displayed in Fig.~\ref{fignsrstrong}, whereas the $d{n_{\rm s}}/dln k - {n_{\rm s}}$ diagram is shown in Fig.~\ref{strongfignsdns}. One can see from these plots, that our model can fit the data from Planck, and then describe the inflationary era.

\begin{table}
  \centering
\begin{tabular}{cccccccccc}
\\
\hline
\hline
$n$ & $\alpha$ & $t_\star$ & $t_{\rm e}$ & $\Gamma_0$ & $\beta=\kappa$ & $n_{\rm s}$ & $r\ast10^{-9}$ & $\epsilon_1^\star$ & $\eta^\star$ \\
\hline\\
$0.02$ & $0.31$ & $11.7939$ & $0.0152652$ & $1.02989$ & $0.18$ & $0.960136$ & $0.90595$ & $0.000868499$ & $0.0986662$ \\
    $0.021$ & $0.32$ & $17.6184$ & $0.0555765$ & $1.02535$ & $0.19$ & $0.966059$ & $1.89603$ & $0.0016062$ & $0.10164$ \\
    $0.022$ & $0.32$ & $19.8509$ & $0.103392$ & $1.04208$ & $0.2$ & $0.944485$ & $2.30056$ & $0.0021761$ & $0.0929556$ \\
\hline
\\
  \end{tabular}
\caption{To estimate the free parameters of the model we considered Planck 2018 data set and used the results of Fig.\ref{StrongPlotNsraa}.}\label{Tab1}
\end{table}


\section{Conclusion}
\label{sec:conclusions}
In this paper, we have used string field theory and then analyse the possibility of explaining inflation within this model. This theory is based on a non-local tachyonic action. The non-locality of the action makes computations complicated, but one can introduce an effective scalar field to rewrite the action in a neat way and then, computations become much simpler. The flat FLRW cosmological equations are employed as the background cosmological equations and after imposing some ansatz for the boundary conditions and the form of the model (power-law form of the cosmic time). Thus, it is possible to write  exact cosmological equations for the Hubble parameter and the potential. We use these for  performing  scalar cosmological perturbations with the aim of finding out the standard inflationary parameters. It was observed that the  perturbation in the background metric could  produce a Mukhanov-Sasaki like equation, and so the  standard cosmological perturbation theory has  be used for this model. Thus, we  calculated the slow-roll parameters for this  non-local cosmological model, and then we also explicitly found the most important inflationary parameters, i.e., the  scalar and tensor power spectrum,  their related indices, and the  tensor-to-scalar ratio. Then, we used cosmological data from Planck data set to constraint the free parameters in this model. The most important results of this papers are summarized in Figs.~\ref{StrongPlotNsraa}-\ref{strongfignsdns} where we constrain our parameters using the recent cosmological data sets and plot the corresponding inflationary parameters. It is then found that this model can explain the current observations coming from the early universe, as an inflationary phase of the universe.

Let us note that the non-locality that we have  investigated is not artificial, as it  comes from the zero level modes of string field theory. It is then interesting to note that inflation can be explained within this well-motivated theoretical model. It would be interesting to further generalise this study with other possible observational constrains that one can use from other situations like for example gravitational waves. One way of doing this would be to take  tensorial perturbations for the FLRW metric and compute the velocity of the propagation of gravitational waves $c_T$, and then constraining this value to the speed of light. This is nowadays one of the most important constrains coming from LIGO~\cite{Goldstein:2017mmi}, and many theories have been partially ruled out due to this constrain. Another important route would be to analyse other properties of gravitational waves such as its polarization or even analyze if is possible to generate wave forms of this model, for example for binary merger of black holes or neutron stars. These studies will be performed in the future.

It may be noted that gravitational waves can be produced from scalar perturbations \cite{gw12, gw14}. The production of such  gravitational waves from   scalar perturbations has also been studied in  a string axion inflationary cosmology \cite{gw15}.
It was demonstrated in such a study that the  non-perturbative effects can produce correction to the  original  axion potential. It was also observed that this model of inflationary cosmology
  matches the Planck observations. This corrected axion potential  also produced  short distance  enhancement in the scalar power spectrum.
  The production of  gravitational wave   during inflation has been studied in  type IIB string theory \cite{gw16}. It has been suggested that it might be possible to observe these gravitational waves.
 It has been demonstrated that chiral gravitational waves can be produced because of the axion-gauge field coupling, in such models. As the production of gravitation waves during inflation has been studied in string theory, it would also be interesting to study the production of gravitational waves due to scalar perturbations in string field theory.

 It may be noted that   heterotic string  field theory has also been constructed \cite{super4, super5}.
 The tree level effective action of the heterotic string has been studied  \cite{super1}. It has been demonstrated that the light fields  are charged under an  $\mathcal{N}=2$ $R$-charge in the left-moving sector.   The  Neveu-Schwarz sector of heterotic string field theory has also been constructed
 \cite{super2}. It has been demonstrated that the   action for the heterotic string field theory  has  a novel  nonpolynomiality, and contains terms terms needed  to cover missing regions of moduli spaces \cite{super20}.
 Inflation has been studied in  heterotic string theory using   non-perturbative effects \cite{super6}.  The     stabilization of modulus associated to the inflaton has also been discussed in heterotic string theory \cite{super7}. Thus, it is possible to study
 inflation in heterotic string theory. As heterotic string field theory has been constructed, it would be interesting to analyze inflation in heterotic string field theory.
It would thus be interesting to find the  the slow-roll parameters for a inflationary model constructed from heterotic string field theory. We can then  obtain the  scalar and  tensorial power spectrum,   their related indices in heterotic string field theory. It would also be interesting to use   cosmological data from Planck 2018 to constrain the free parameters in this model of inflationary cosmology.

It may be noted that as the  T-duality is an important symmetry in string theory, it is possible to define a generalized geometry based on string theory, which is manifestly invariant under T-duality \cite{dual01, dual02, dual1,  dual2}. It has been possible to construct models of cosmology based on this generalized geometry  \cite{dual4, dual5, dual12, dual14}. It has also been possible to obtain $\alpha'$ correction to these cosmological models using double field theory \cite{dual6}. It is also possible to analyze the T-duality of string field theory \cite{dual7}.  It would be interesting to construct a cosmological model using T-duality of string theory. This cosmological model can then be used to study inflation. In fact, it would be interesting to analyze inflationary cosmology using the generalized geometry, with manifest T-duality.
Thus, it would be interesting to construct  the slow-roll parameters for a inflationary model in generalized geometry.  It is expected that the manifest T-duality of such a geometry will produce non-trivial effects for inflationary cosmology.  So, it  is   expected that the manifest T-duality of generalized geometry can have non-trivial consequences for the scalar and  tensorial power spectrum of inflationary cosmology.
As string theory can produce interesting results for  the production of gravitational waves during inflation
\cite{gw15, gw16}, it would be interesting to analyze the production of such gravitational waves using generalized geometry.

\section*{Acknowledgements}
H.S. thanks A. Starobinsky for very constructive discussions about inflation during Helmholtz International Summer School  2019 in Russia. H.S. also thanks G. Ellis, A. Weltman, and UCT for arranging his short visit, and for enlightening discussions about cosmological fluctuations and perturbations for both large and local scales. H.S. is also graceful to T. Harko and H. Firouzjahi for constructive discussions about inflation and perturbations. Finally, H.S. gives as special thanks to his wife, E. Avirdi, for her patience during their stay in South Africa. S.B. is supported by Mobilitas Pluss  MOBJD423 by the Estonian government.

\end{document}